\documentclass[journal]{IEEEtran}
\usepackage{amsmath,amsfonts}
\usepackage{array}
\usepackage{textcomp}
\usepackage{stfloats}
\usepackage{url}
\usepackage{verbatim}
\usepackage{graphicx}
\usepackage{balance}
\usepackage{cite}
\usepackage{cleveref}
\usepackage{dsfont}
\usepackage{color}
\usepackage[caption=false,font=footnotesize]{subfig}
\usepackage{mathtools, cuted}
\usepackage[export]{adjustbox}

\usepackage{microtype}
\usepackage{multicol}
\usepackage{booktabs}

\usepackage{amsmath}
\usepackage{amssymb}
\usepackage{mathtools}
\usepackage{amsthm}

\newcommand{\Avg}{{\textrm{Avg}}}
\newcommand{\ROI}{{\textrm{ROI}}}
\newcommand{\ROP}{{\textrm{ROP}}}
\newcommand{\RONI}{{\textrm{RONI}}}
\newcommand{\CPP}{{\textrm{CPP}}}
\newcommand{\PSNR}{{\textrm{PSNR}}}
\newcommand{\SSIM}{{\textrm{SSIM}}}
\newcommand{\SNR}{{\textrm{SNR}}}
\newcommand{\MSE}{{\textrm{MSE}}}
\newcommand{\MAXX}{{\textrm{MAX}}}
\newcommand{\inn}{{\textrm{in}}}

\newcommand{\norm}[1]{\left\lVert#1\right\rVert}
\begin{document}

\title{Region-of-Interest-Guided Deep Joint Source-Channel Coding for Image Transmission}
\author{
  Hansung Choi and Daewon Seo
  \thanks{The authors are with the Department of Electrical Engineering and Computer Science, Daegu Gyeongbuk Institute of Science and Technology (DGIST), Daegu 42988, South Korea (e-mail: \{hansungchoi, dwseo\}@dgist.ac.kr).}
}

\maketitle

\begin{abstract}
Deep joint source-channel coding (deepJSCC) methods have shown promising improvements in communication performance over wireless networks. However, existing approaches primarily focus on enhancing overall image reconstruction quality, which may not fully align with user experiences, often driven by the quality of regions of interest (ROI). Motivated by this, we propose ROI-guided joint source-channel coding (ROI-JSCC), a novel deepJSCC framework that prioritizes high-quality transmission of ROI. The ROI-JSCC consists of four key components: (1) Image ROI embedding, (2) ROI-guided split processing, (3) ROI-based loss function design, and (4) ROI-adaptive bandwidth allocation. Together, these components enable ROI-JSCC to selectively enhance the ROI reconstruction quality at varying ROI positions while maintaining overall image quality with minimal computational overhead. Experimental results under diverse communication environments demonstrate that ROI-JSCC significantly improves ROI reconstruction quality while maintaining competitive average image quality compared to recent state-of-the-art methods.
\end{abstract}

\begin{IEEEkeywords}
joint source-channel coding, region-of-interest, image transmission
\end{IEEEkeywords}

\section{Introduction} \label{sec:introduction}
With the arrival of the 6G era, a wide range of applications is anticipated, including autonomous driving, smart factories, digital twins, and extended reality technologies such as virtual and augmented reality (VR/AR)~\cite{getu2025semantic}. These applications place immense demands on wireless networks, particularly in terms of reliability and latency. However, conventional digital communication systems---refined over decades to ensure bit-level accuracy---struggle to meet these requirements. Further improvements to conventional systems have become increasingly difficult, posing significant challenges in supporting the demanding transmission needs of next-generation applications.

To address these challenges, deep learning-based joint source-channel coding (deepJSCC)~\cite{bourtsoulatze2019deep,yang2024swinjscc} and semantic communication (SemCom)~\cite{yu2025multi} systems have been proposed. DeepJSCC models jointly learn the characteristics of source data and communication channels, enabling the mapping of source data directly into robust and informative channel input symbols. By leveraging the capability of deep learning models, these models have shown high performance in preserving data quality. In contrast, task-oriented SemCom focuses on transmitting task-relevant information, thereby reducing communication overhead while sustaining task performance in applications such as multi-task learning~\cite{yu2025multi}. More recently, region-of-interest-guided semantic communication (ROI-SemCom) methods are proposed to enhance task performance further~\cite{wang2025task,tan2023learned}. These methods identify task-determined ROI areas and allocate more bandwidth to them while assigning less to other areas, thus improving task performance.

However, these methods often fail to account for an essential requirement in real-world applications, namely, the need to reconstruct both the important region and the whole image with sufficient accuracy. For instance, current autonomous driving systems can avoid sudden obstacles using collision prediction, but emergency braking of a leading vehicle often causes rear-end collisions due to insufficient reaction time. Since such accidents dominate autonomous vehicle crashes~\cite{national2022summary}, sharing the scenes where the obstacles are presented more clearly is essential for timely response of following vehicles. In this case, the ROI area containing the suddenly appearing object predicted as high collision probability should be delivered with high accuracy while preserving overall image quality to better support the following car’s decision-making. Although the aforementioned current deepJSCC, SemCom, and ROI-SemCom have achieved remarkable success in overall performance, they fail to meet this goal since they treat the suddenly appearing obstacle and a less relevant distant object in the same way. Moreover, such approaches require retraining or fine-tuning when the ROI position varies.

To address the above issue, we propose a region-of-interest-guided deepJSCC (ROI-JSCC). Our ROI-JSCC focuses on the ROI of images themselves and adaptively enhances the transmission quality of any given image ROIs while preserving average image quality with the following four mechanisms.

\noindent \textbf{(1) Image ROI embedding:} This mechanism integrates ROI information into intermediate deepJSCC features, enabling the model to enhance ROI areas. Unlike previous ROI-SemCom embedding methods~\cite{wang2025task,tan2023learned}, which rely on task-determined ROIs and cannot handle ROI areas that differ from the task definition, our method can dynamically accommodate varying image ROIs without retraining or fine-tuning.

\noindent \textbf{(2) ROI-guided split processing:}  Important ROI features are refined using intensive self-attention~\cite{yang2024swinjscc}, while less important features are processed with lightweight spatial attention~\cite{woo2018cbam}. Unlike previous deepJSCC that applies intensive self-attention for all features without considering ROI importance~\cite{yang2024swinjscc}, ours is the first computational mechanism that applies self-attention to the only ROI related features. This efficiently enhances the processing capability for ROI features with low computational burden.

\noindent \textbf{(3) ROI-based loss function design:} Leveraging the mechanism in (1), a new loss function guides ROI-JSCC to learn which features are related to ROI and how to encode them to preserve their quality in the transmitted image. Unlike previous ROI-SemCom loss functions, which are designed to enhance task-determined ROIs, our loss function enables the model to enhance dynamically varying image ROIs without retraining.

\noindent \textbf{(4) ROI-adaptive bandwidth allocation:} This mechanism significantly increases bandwidth allocation for ROI features while slightly reducing it for others, thereby enhancing ROI quality with minimal degradation elsewhere. The previous ROI-SemCom methods~\cite{wang2025task,tan2023learned} need additional transmission to inform the bandwidth allocation result of every feature pixel. In contrast, our method only needs to transmit a few bits of the ROI's coordinate location, which significantly reduces the extra transmission burden.

These mechanisms enable ROI-JSCC to improve ROI-specific image quality across varying ROI configurations without requiring retraining or fine-tuning. Moreover, it efficiently utilizes computational resources by allocating more resources to the ROI area.

\begin{figure*}[t]
    \centering
    \includegraphics[width=1.0\linewidth]{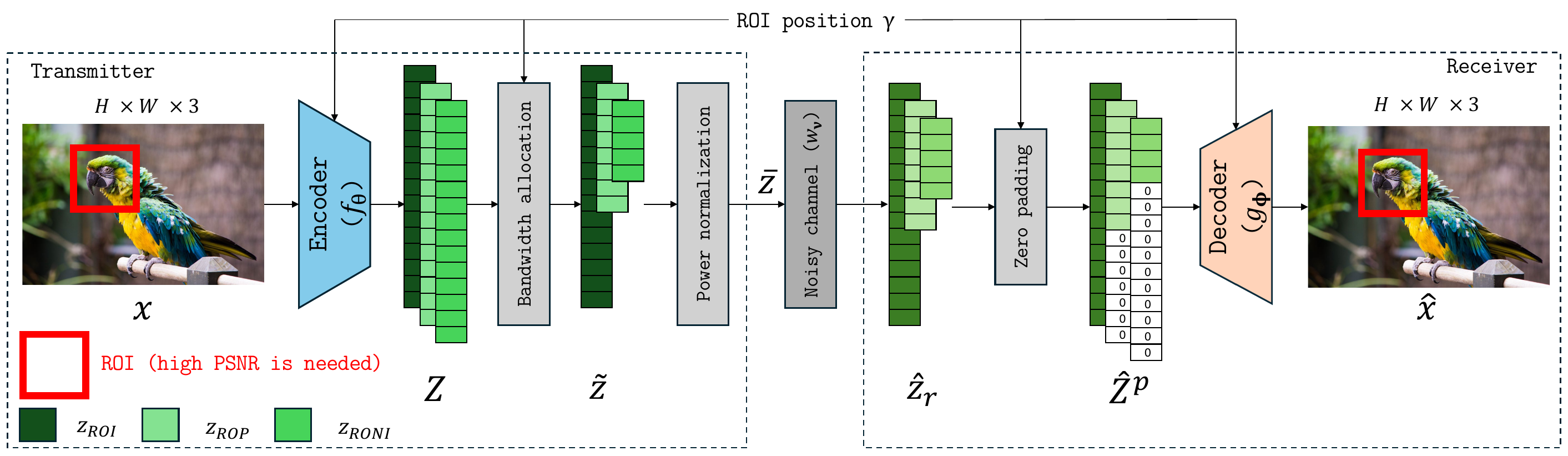}
    \caption{ROI-JSCC communication system.}\label{fig:ROIJSCC_system}
\end{figure*}

\section{ROI-JSCC Communication System} \label{sec:ROI-JSCC_Overview}
We first partition the image into $n_h \times n_w$ patches, among which our primary focus is a single patch representing the region-of-interest (ROI). We assume that the position of ROI $\gamma := (h_\gamma, w_\gamma)$ in the image is known to the transmitter and receiver, where $h_\gamma$ and $w_\gamma$ denote the vertical and horizontal indices in the ROI map, see Figure~\ref{fig:mask_generation}.\footnote{In the driving example in Section~\ref{sec:introduction}, the transmitter identifies the obstacle using an algorithm, e.g., segmentation, and determines the ROI's location with obstacle's collision probability. This ROI position needs to be shared with the receiver; however, since it requires only $\log_2 (n_h n_w)$ bits, negligible relative to the total bandwidth usage, we ignore this communication overhead. For example, when $n_h=n_w=4$, an additional $4$ bits transmission is required. Thus, we can assume that the ROI position is available to both the transmitter and the receiver. More detailed explanations are provided in Appendix~\ref{app:scenario}.} Adjacent patches are categorized as region of periphery (ROP) or region of non-interest (RONI) according to their relative location with respect to the ROI, see Figure~\ref{fig:mask_generation} for example. Our objective is to transmit images in a manner that preserves high quality in ROI, which dominates user experience, while minimizing degradation in overall image quality at the same time. To this end, we propose a ROI-JSCC communication system illustrated in Figure~\ref{fig:ROIJSCC_system}.

In this system, the encoder $f_{\theta}$ extracts a feature matrix $\mathbf{Z} := [\mathbf{z}_1, \ldots, \mathbf{z}_B]^T \in \mathbb{C}^{B \times C_m}$ from a source image $\mathbf{x} \in \mathbb{R}^{H \times W \times 3}$ using the ROI position $\gamma$, where $H$, $W$, and $3$ denote the height, width, and RGB channels of $\mathbf{x}$, respectively. Among the features, some contain significantly more information about the ROI; we refer to these as ROI feature vectors, denoted by $\mathbf{z}_{\ROI}$. To preserve the information in $\mathbf{z}_{\ROI}$, the transmitter allocates greater bandwidth to $\mathbf{z}_{\ROI}$ and less bandwidth to the remaining features by transforming each feature vector dimension, under a total bandwidth constraint. The feature vectors are then aggregated into $\mathbf{\tilde{z}} \in \mathbb{C}^k$, where $k$ denotes the total channel bandwidth usage. The channel usage per pixel (CPP) is defined as the ratio of the channel bandwidth to the total number of RGB pixels in $\mathbf{x}$, i.e., $\CPP := \frac{k}{3HW}$. Next, the transmitter rescales $\mathbf{\tilde{z}}$ into $\mathbf{\bar{z}}$ to fit the power constraint $P$. Throughout this paper, we assume $P=1$, i.e., $\frac{1}{k} \norm{\mathbf{\bar{z}}}_{2}^{2} \le 1$.

After power normalization, $\mathbf{\bar{z}}$ is transmitted over a wireless noisy channel. The receiver then obtains the corrupted feature vector $\mathbf{\hat{z}}_r \in \mathbb{C}^{k}$. In this paper, we consider an additive white Gaussian noise (AWGN) and a fast Rayleigh fading channel, modeled as
$\hat{z}_{r,i} = h_i \bar{z}_i + \epsilon_i, ~~ \mathbf{\epsilon_i} \sim \mathcal{CN}(0,\sigma^2)
$
where $\mathcal{CN}(0, \sigma^2)$ denotes a complex Gaussian distribution and subscript $i$ denotes $i$-th symbol of the corresponding vector. Here, $h_i=1$ for the AWGN channel and $h_i \sim \mathcal{CN}(0,1)$ for the fast Rayleigh fading channel. The quality of the communication channels is characterized by the signal-to-noise ratio (SNR), defined as
$
\SNR := 10 \log_{10} \left( 1/\sigma^2 \right)~ \text{dB}. 
$

At the receiver, the received vector $\mathbf{\hat{z}}_r \in \mathbb{C}^{k}$ is recovered into $\mathbf{\hat{Z}}^p := [\mathbf{\hat{z}}^p_1, \ldots, \mathbf{\hat{z}}^p_B]^T \in \mathbb{C}^{B \times C_m}$ using zero-padding based on the ROI position $\gamma$, as illustrated in Figure~\ref{fig:ROIJSCC_system}. The decoder $g_{\phi}$ then reconstructs an image $\mathbf{\hat{x}}$ from $\mathbf{\hat{Z}}^p$ and $\gamma$. The quality of the reconstructed image $\mathbf{\hat{x}}$ is evaluated using the peak signal-to-noise ratio (PSNR) and the structural similarity index measure (SSIM). The PSNR quantifies pixel-wise distortion:
$
\PSNR := 10 \log_{10} \frac{255^2}{\MSE (\mathbf{x}, \mathbf{\hat{x}})} ~ \text{dB},
$
and the SSIM is a perceptual metric:
$
\SSIM(\mathbf{x},\hat{\mathbf{x}}) := l(\mathbf{x},\hat{\mathbf{x}})^{p} c(\mathbf{x},\hat{\mathbf{x}})^{q} s(\mathbf{x},\hat{\mathbf{x}})^{r},
$
where $l(\cdot)$, $c(\cdot)$, and $s(\cdot)$ measure brightness, contrast, and structural similarities and $p, q, r$ weight contributions of each measure. In this paper, our objective is to improve the performance in the ROI ($\PSNR_{\ROI}$, $\SSIM_{\ROI}$) while maintaining comparable average performance ($\PSNR_{\Avg}$, $\SSIM_{\Avg}$) with respect to previous deepJSCC models. In the following subsections, we introduce the encoder and decoder architectures of ROI-JSCC, the ROI-based loss function, and present the detailed procedure for ROI-adaptive bandwidth allocation.

\subsection{Encoder and Decoder Design}

\begin{figure}[t]
    \centering
    \includegraphics[width=0.8\linewidth]{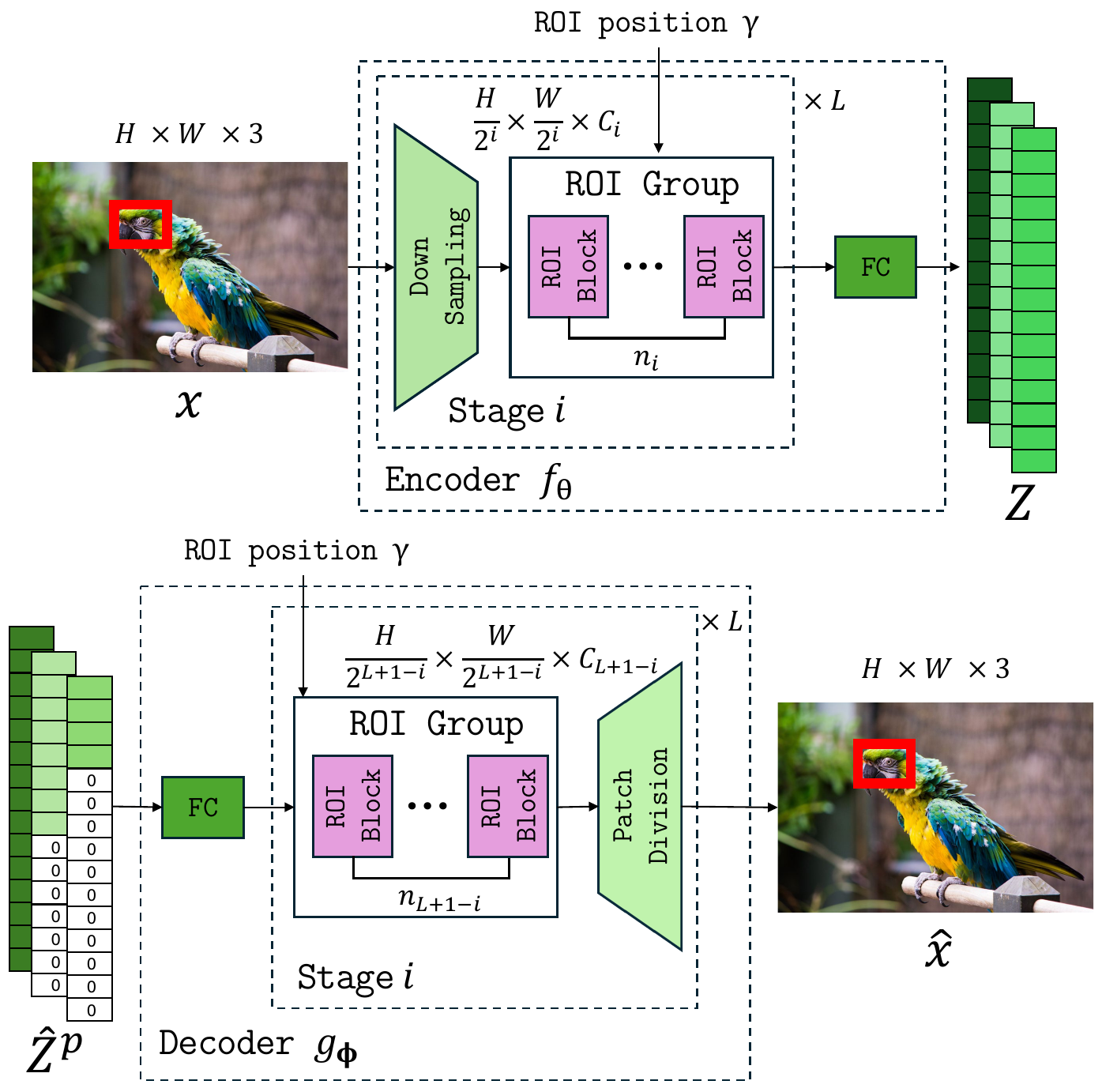}
    \caption{The structures of the encoder and the decoder.}\label{fig:Encoder_Decoder}
\end{figure}

The architectures of the encoder $f_{\theta}$ and decoder $g_{\phi}$ are illustrated in Figure~\ref{fig:Encoder_Decoder}. Both consist of $L$ stages. For the encoder, at stage $i$, a downsampling layer reduces the feature resolution from $\frac{H}{2^{i-1}} \times \frac{W}{2^{i-1}} \times C_{i-1}$ to $\frac{H}{2^{i}} \times \frac{W}{2^{i}} \times C_i$. In our model, a patch embedding layer is used at stage 1, while patch merging layers~\cite{yang2024swinjscc} are employed for the subsequent stages. Each stage then processes features using a region-of-interest-guided group (ROI group), which consists of $n_i$ region-of-interest-guided blocks (ROI blocks). Leveraging the ROI position $\gamma$, these ROI blocks prioritize enhancing features relevant to the ROI while still refining others. Finally, a fully connected (FC) layer maps each feature dimension from $C_L$ to $2C_M$. Through this process, the encoder extracts the semantic feature matrix $\mathbf{Z} \in \mathbb{C}^{B \times C_m}$ from the input image $\mathbf{x}$, where $B := \frac{H}{2^L} \times \frac{W}{2^L}$.

For the decoder, an FC layer first transforms each feature dimension from $2C_M$ back to $C_L$. Then, for each stage $i$, an ROI group processes the features using $\gamma$, followed by a patch division layer~\cite{yang2024swinjscc} that upsamples the features from $\frac{H}{2^{L+1-i}} \times \frac{W}{2^{L+1-i}} \times C_{L+1-i}$ to $\frac{H}{2^{L-i}} \times \frac{W}{2^{L-i}} \times C_{L-i}$. Through this decoding pipeline, the decoder reconstructs the image $\mathbf{\hat{x}}$ from the input feature matrix $\mathbf{\hat{Z}}^p$. In the following subsection, we describe the internal structure of the ROI block in detail.

\subsection{ROI Block Details}

\begin{figure}[t]
    \centering
    \includegraphics[scale=0.45]{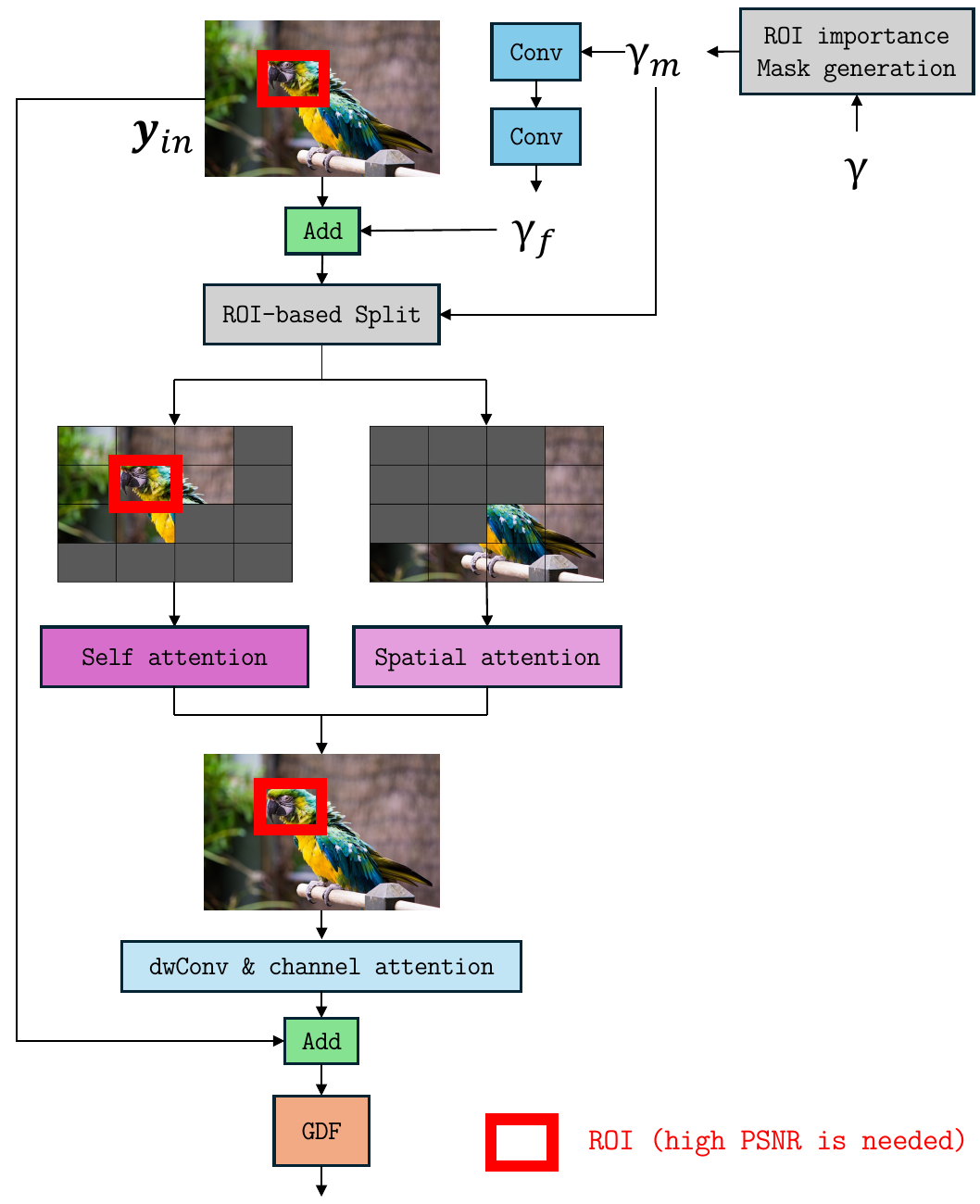}
    \caption{ROI block structure and process for ROI $\gamma =(2,2)$.}\label{fig:ROI_block}
\end{figure}

\begin{figure}[t]
    \centering
    \includegraphics[scale=0.35]{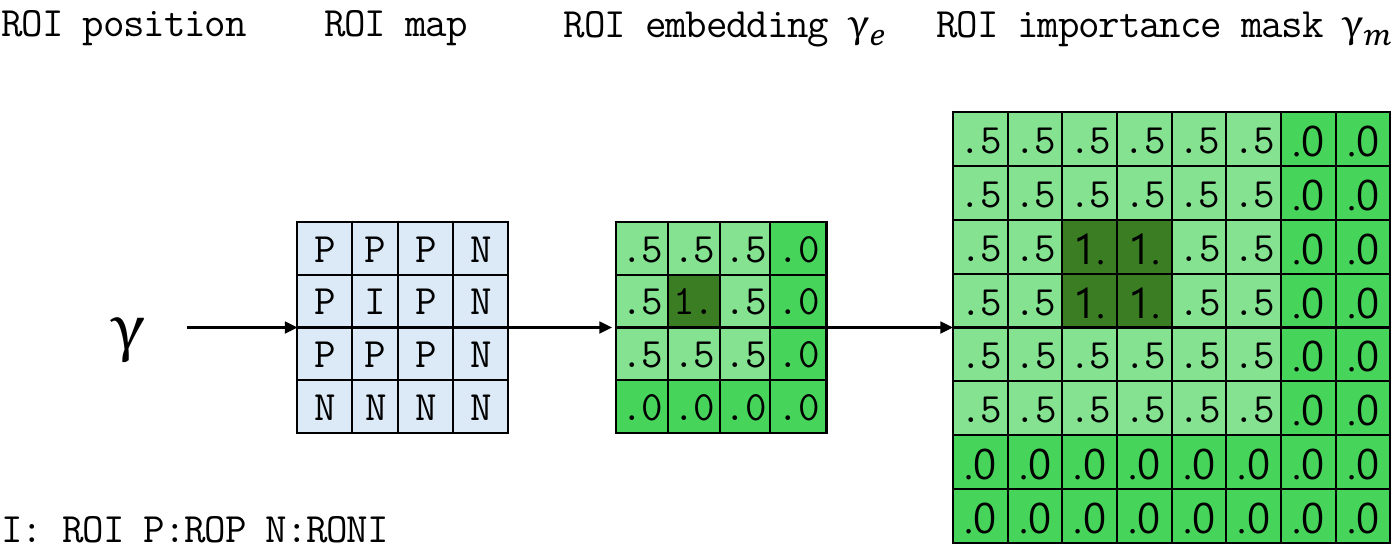}
    \caption{The ROI importance mask generation process for ROI $\gamma =(2,2)$.}\label{fig:mask_generation}
\end{figure}

The detailed structure of the proposed ROI block is illustrated in Figure~\ref{fig:ROI_block}.\footnote{To illustrate, the refinement process is visualized using an image, but the ROI block in fact operates on intermediate neural features rather than raw pixel data.} First, the ROI importance mask $\gamma_m$ is obtained through the generation process depicted in Figure~\ref{fig:mask_generation}. Based on the ROI position $\gamma$, the ROI map is constructed by labeling patches adjacent to the ROI as ROP and all other patches as RONI. Next, the ROI embedding $\gamma_e$ is produced by assigning the values $1.0$, $0.5$, and $0.0$ to the ROI, ROP, and RONI patches, respectively. The ROI importance mask $\gamma_m$ is then generated by applying the nearest-neighbor method to upsample $\gamma_e$ to match the resolution of the ROI block’s input feature $\mathbf{y}_{\inn}$. To effectively incorporate ROI information into $\mathbf{y}_{\inn}$ without degrading its original content, $\gamma_m$ is further converted into an importance feature map $\gamma_f$, which is then added to $\mathbf{y}_{\inn}$. This integration mechanism enables subsequent layers to identify and emphasize features that are most relevant for improving $\PSNR_\ROI$.

Then, the feature vectors in $\mathbf{y}_{\inn}$ are categorized based on the values in the ROI importance mask $\gamma_m$: features corresponding to values of $1$, $0.5$, and $0.0$ are defined as ROI feature vectors $\mathbf{z_\ROI}$, ROP feature vectors $\mathbf{z_\ROP}$, and RONI feature vectors $\mathbf{z_\RONI}$, respectively. To enhance the processing capability of $\mathbf{z_\ROI}$ without increasing computational cost, the ROI block applies separate processing strategies to $\mathbf{z_\ROI}$, $\mathbf{z_\ROP}$, and $\mathbf{z_\RONI}$.

As illustrated in Figure~\ref{fig:ROI_block}, $\mathbf{z_\ROI}$ and the majority of $\mathbf{z_\ROP}$ are processed using computationally-intensive self-attention~\cite{yang2024swinjscc}, whereas $\mathbf{z_\RONI}$ and the $\mathbf{z_\ROP}$ vectors adjacent to the three RONI are processed using lightweight spatial attention~\cite{woo2018cbam}. Self-attention has proven to be a powerful tool for feature extraction in deepJSCC~\cite{yang2024swinjscc}, whereas spatial attention provides satisfactory performance with low computational complexity. By leveraging both mechanisms under ROI guidance, the proposed split processing not only enhances $\PSNR_{\ROI}$ but also minimizes degradation in $\PSNR_{\Avg}$, which is the first efficient mechanism that allocates different computational resources to ROI and non-ROI regions. 

In communication systems, enhancing feature correlation contributes to error correction in channel coding, while reducing redundant information is essential for efficient source coding. To achieve both objectives, the proposed ROI block processes the entire set of features jointly by applying depthwise convolution and channel attention, thereby strengthening feature correlations. In addition, it employs a gated depthwise convolutional feed-forward network (GDF)~\cite{zamir2022restormer} to suppress redundancy and improve feature compactness.

\subsection{ROI-based Loss Function}
Deep neural models have the capacity to learn the relationship between conditional inputs and the loss function. In our setting, the conditional input is $\gamma$, and our goal is to maximize the image quality of the ROI region while minimizing the average performance degradation. To train ROI-JSCC for improving $\PSNR_{\ROI}$ across varying $\gamma$, we design the ROI-based loss function $\mathcal{L}$ as follows:
$
   \mathcal{L}(\mathbf{x},\mathbf{\hat{x}}) := d(\mathbf{x},\mathbf{\hat{x}}) + \alpha d(\mathbf{x_{\ROI}},\mathbf{\hat{x}_{\ROI}}) + \beta d(\mathbf{x_{\ROP}},\mathbf{\hat{x}_{\ROP}}),
$
where $0 < \beta < \alpha < 1$ are hyperparameters and $d(\cdot)$ is the mean squared error (MSE) for the PSNR metric and $1-\SSIM$ for the SSIM metric. By using this loss function, ROI-JSCC learns to prioritize features that contribute to improving the ROI region quality with the aid of $\gamma$, while preserving average quality. Consequently, it effectively maintains high quality in the ROI during image transmission under varying $\gamma$ without fine-tuning.

\subsection{ROI-based Bandwidth Allocation}
The possible average transmittable dimension per feature vector $C_{\Avg}$ under the bandwidth constraint $k$ can be calculated by the following equation: $B \cdot C_{\Avg} = k$, where $B$ is the total number of feature vectors. When the image is divided into $n_h \times n_w$ patches, the number of ROI patches $n_\ROI$, ROP patches $n_\ROP$, and RONI patches $n_\RONI$ satisfy the following patch number inequalities: $n_\ROP \le 8 n_\ROI$, $n_\RONI \le n_h n_w - 9 n_\ROI$ since each ROI patch can have at most $8$ adjacent ROP patches. Let $C_\ROI$ and $C_\RONI$ be the increased and decreased dimension for ROI and RONI feature vectors. Then, the bandwidth constraint $k$ is satisfied if the next inequality 
$
n_\ROI \cdot C_{\ROI} \le n_\RONI \cdot C_\RONI \le (n_h n_w - 9 n_\ROI) C_\RONI,
$
holds by the patch number inequalities. Therefore, with $\eta=(n_h n_w - 9 n_\ROI)/n_\ROI$, assigning dimensions of $(1 + \eta \tau) C_{\Avg}$, $C_{\Avg}$, and $(1 - \tau) C_{\Avg}$ to $\mathbf{z_\ROI}$, $\mathbf{z_\ROP}$, and $\mathbf{z_\RONI}$, respectively, satisfies the bandwidth constraint $k$, where $0 < \tau < 1$ is a hyperparameter. This ROI-based bandwidth allocation significantly increases the bandwidth assigned to $\mathbf{z_\ROI}$ while marginally reducing that for $\mathbf{z_\RONI}$. As a result, the ROI quality is enhanced with only a minimal loss in other regions.

\begin{figure*}[t]
\centering
\begin{tabular}{@{}p{0.64\linewidth}@{\hspace{0.3em}}c@{\hspace{0.3em}}p{0.24\linewidth}@{}}
\centering
\includegraphics[width=1.0\linewidth]{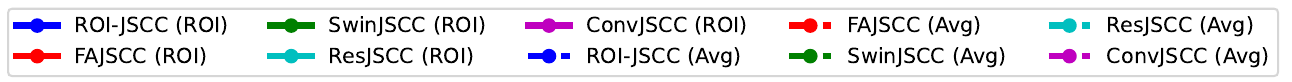}
&
\vrule width 0.5pt
&
\includegraphics[width=0.85\linewidth]{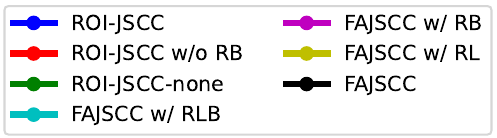}
\\[1em]
\begin{minipage}{\linewidth}
\centering
\includegraphics[width=0.33\linewidth]{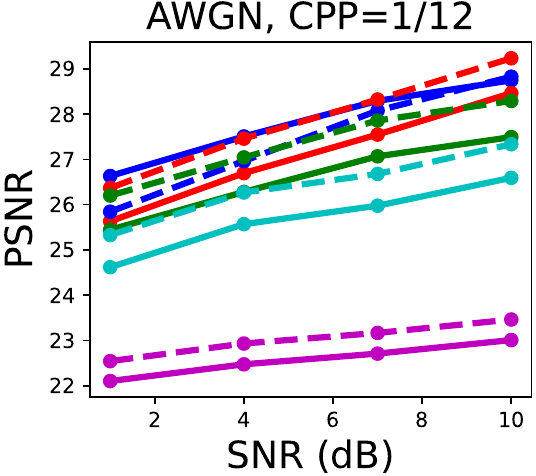}\hfill
\includegraphics[width=0.33\linewidth]{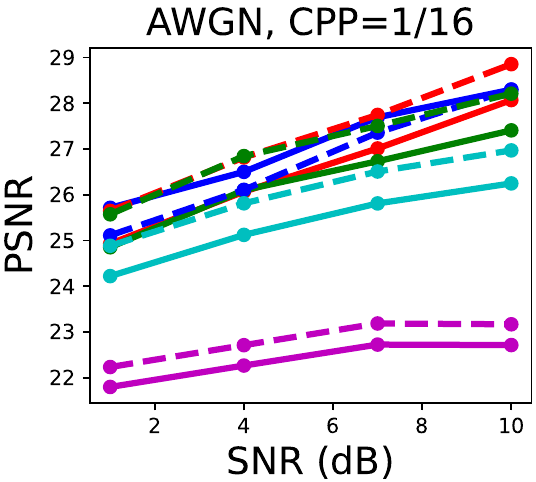}\hfill
\includegraphics[width=0.33\linewidth]{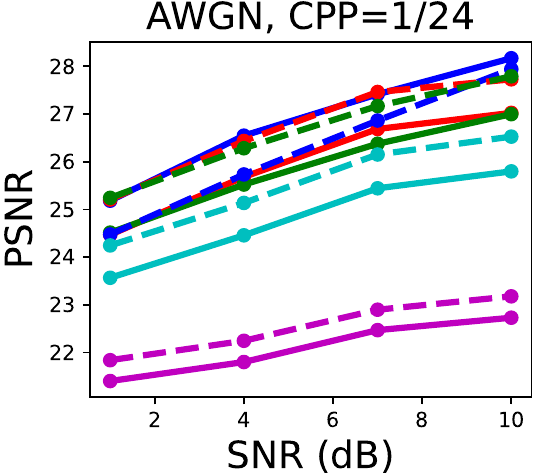}
\end{minipage}
&
\vrule width 0.5pt
&
\begin{minipage}{\linewidth}
\centering
\includegraphics[width=0.8\linewidth]{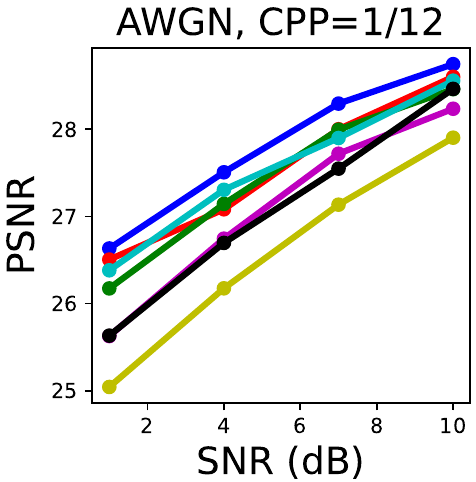}
\end{minipage}
\end{tabular}

\caption{The left figures of the vertical line show $\PSNR_{\ROI}$ (solid lines) and $\PSNR_{\Avg}$ (dotted lines) results of DIV2K validation dataset under different CPP and SNR environments. The right figure of the vertical line shows the $\PSNR_{\ROI}$ results of the DIV2K validation dataset for the ablation study of the ROI-based mechanisms in ROI-JSCC.}
\label{fig:main_result}
\end{figure*}

\begin{figure*}
\centering
\includegraphics[width=0.65\linewidth]{Figure/plot_legend_main2.pdf}\\
\begin{multicols}{4}
\centering
\includegraphics[width=1.0\linewidth]{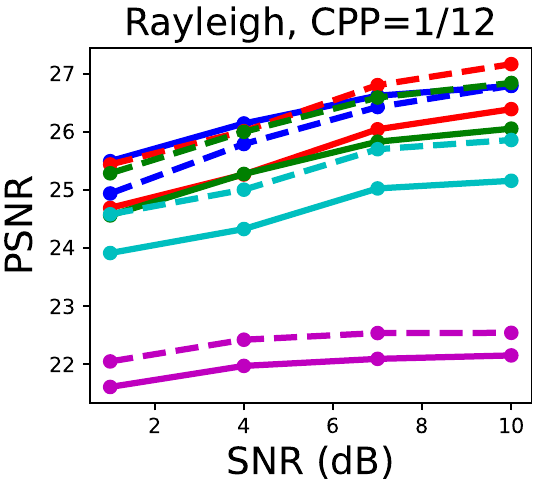}\\

\includegraphics[width=1.0\linewidth]{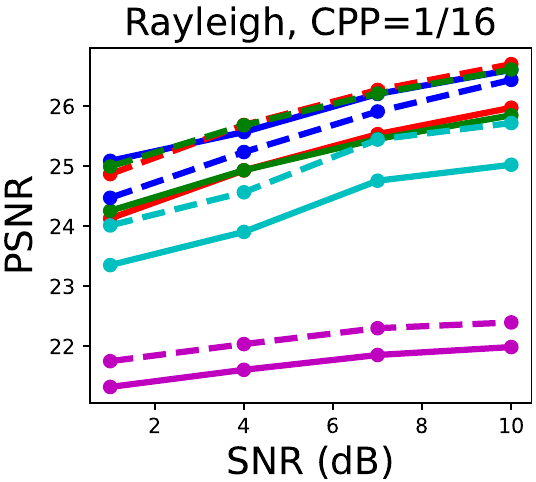}\\

\includegraphics[width=1.0\linewidth]{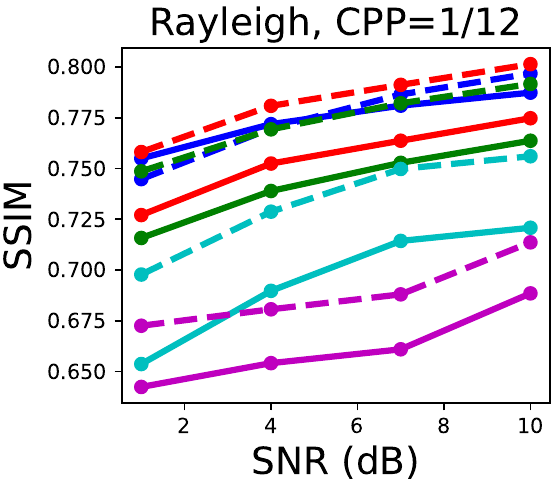}\\

\includegraphics[width=1.0\linewidth]{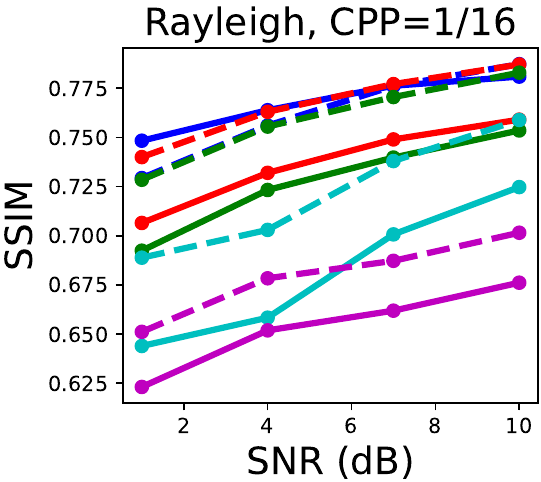}\\
\end{multicols}

\caption{ROI and average performance results under a fast Rayleigh fading channel. The left two figures show PSNR results, and the right two figures show SSIM results. Solid lines are for ROI performances and the dotted lines are for average performances.}\label{fig:Rayleigh_result}
\end{figure*}

\section{Experiment}
\label{sec:experiment}

\subsection{Experimental Setting}
\label{sec:experiment_setting}
\noindent \textbf{Baselines:} We compare ROI-JSCC with previous deepJSCC models ranging from the original~\cite{bourtsoulatze2019deep} to the latest~\cite{choi2025feature} to verify our model. The baselines include a convolution block-based ConvJSCC~\cite{bourtsoulatze2019deep}, a residual block-based ResJSCC~\cite{zhang2023predictive}, a Swin transformer block-based SwinJSCC~\cite{yang2024swinjscc}, and a feature importance-aware block-based FAJSCC~\cite{choi2025feature}. Note that ours not only improves the ROI performance but also adapts to dynamic positional variations of ROI.

\noindent \textbf{Training and Testing:}
We use the DIV2K dataset for training. The training set consists of 800 high-resolution images, randomly cropped into $256 \times 256$ patches with a batch size of $11$ over $200$ epochs. The Adam optimizer with a learning rate of $0.0001$ is used to minimize the ROI-based loss function $\mathcal{L}$ (with $n_h=n_w=4$, varying $\gamma$, $\alpha=1.0$, $\beta=0.5$, and $\tau=0.1$) for ROI-JSCC, while other models are trained to minimize the MSE loss. For numerical evaluation, varying-sized images of the DIV2K validation dataset are center-cropped to the nearest multiple of $128$ to match the original size closely. All codes for experiments are available at https://github.com/hansung-choi/ROI-JSCC.\footnote{Our code repository includes implementations for additional experiments (e.g., visual inspections and various ROI settings). More detailed explanations are provided in Appendix~\ref{app:extensive}.}

\subsection{Experimental Result}
In Figure~\ref{fig:main_result}, the three left figures of the vertical line present the $\PSNR_{\ROI}$ (solid lines) and $\PSNR_{\Avg}$ (dotted lines) results of DIV2K validation dataset under the AWGN channel. In Figure~\ref{fig:Rayleigh_result}, the two left figures show $\PSNR_{\ROI}$, and $\PSNR_{\Avg}$, and the two right figures show $\SSIM_{\ROI}$, and $\SSIM_{\Avg}$, all under the fast Rayleigh fading channel. During testing, the position of ROI $\gamma$ is randomly sampled from $1 < h_{\gamma}, w_{\gamma} < 4$. As shown, our ROI-JSCC consistently achieves substantial improvements in $\PSNR_{\ROI}$, and $\SSIM_{\ROI}$ across all CPP and SNR conditions, outperforming other baselines. Moreover, for $512 \times 768$ resolution images, our ROI-JSCC uses fewer computational resources ($14.13$ GFLOPs) than the recent two SOTA models, FAJSCC ($14.57$ GFLOPs) and SwinJSCC ($20.04$ GFLOPs). This efficiency is attributed to the ROI-based split processing, which enhances $\PSNR_{\ROI}$ without incurring additional computational overhead. These improvements can be attributed to the four aforementioned ROI-based mechanisms, which effectively boost $\PSNR_{\ROI}$, $\SSIM_{\ROI}$ while minimizing the impact on $\PSNR_{\Avg}$, $\SSIM_{\Avg}$.

Regarding $\PSNR_{\Avg}$, ROI-JSCC significantly outperforms ConvJSCC and ResJSCC, despite its focus on improving $\PSNR_{\ROI}$. While $\PSNR_{\Avg}$ of ROI-JSCC underperforms compared to recent state-of-the-art (SOTA) FAJSCC and SwinJSCC at low SNR, it matches or surpasses them at high SNR. This is because sacrificing more $\PSNR_{\Avg}$ at low SNR is necessary to improve $\PSNR_{\ROI}$, given that enhancing $\PSNR_{\ROI}$ under low SNR conditions is more challenging than at high SNR. Although this is a fundamental limitation, these issues can be addressed by transmitting additional refinement information to increase low $\PSNR$ values of specific patches, which we leave as future work.

\subsection{Ablation Study}
We conduct an ablation study to verify the effectiveness of the four ROI-based mechanisms in ROI-JSCC. In Figure~\ref{fig:main_result}, the right figure of the vertical line shows the $\PSNR_{\ROI}$ results for different configurations where some of the four ROI-based mechanisms in ROI-JSCC are added or removed. In this figure, ``w/o RB,'' ``w/RLB,'' ``w/RB,'' and ``w/RL'' denote without ROI-based bandwidth allocation, with ROI-based loss and bandwidth allocation, with ROI-based bandwidth allocation, and with ROI-based loss, respectively. Using only the ROI-based loss (FAJSCC w/RL) or only the ROI-based bandwidth allocation (FAJSCC w/RB) does not lead to improvements in $\PSNR_{\ROI}$. For w/RL, the ROI-based loss alone confuses FAJSCC since it cannot identify which features are related to $\PSNR_{\ROI}$. Similarly, for w/RB, the ROI-based bandwidth allocation also confuses FAJSCC because it does not understand why certain features are allocated different bandwidths. This indicates that simply applying these vanilla extensions is insufficient to increase $\PSNR_{\ROI}$.

When ROI-based loss and ROI-based bandwidth allocation are applied simultaneously (FAJSCC w/RLB), $\PSNR_{\ROI}$ improves because the decoder can indirectly infer which features relate to $\PSNR_{\ROI}$ from the varying bandwidth allocations. However, FAJSCC w/RLB still underperforms compared to all ROI-JSCC and most ROI-JSCC variants without ROI-based bandwidth allocation. This highlights the effectiveness of our proposed ROI feature map $\gamma_f$ extraction mechanism, which directly incorporates ROI information into subsequent layers. Moreover, comparing ROI-JSCC and ROI-JSCC w/o RB demonstrates that ROI-based bandwidth allocation significantly boosts $\PSNR_{\ROI}$ when combined with an appropriate ROI feature map extraction method (noting that FAJSCC w/RB alone does not significantly increase $\PSNR_{\ROI}$).

To further verify the efficiency of ROI-guided split processing that only applies intensive self-attention to ROI-related features, we compare our ROI-JSCC with two variants. The first one is ROI-JSCC-none, which does not apply self-attention at all. Our ROI-JSCC uses an additional $1.95$ GFLOPs and improves $\PSNR_{\ROI}$ significantly compared to ROI-JSCC-none as shown in Figure~\ref{fig:main_result}. The second is ROI-JSCC-all, which applies self-attention to all features. Although this ROI-JSCC-all uses an additional $4.38$ GFLOPs, $\PSNR_{\ROI}$ improvement is marginal $\sim0.01$ dB. We omitted this curve as it is not clearly discernible.

\section{Conclusion}
In this paper, we propose a region-of-interest-guided deep joint source-channel coding (ROI-JSCC) method. Compared to deepJSCC models ranging from the original to the most recent SOTA, our ROI-JSCC adaptively preserves high quality in the region of interest (ROI), which is strongly correlated with user experience. This is achieved through our proposed $4$ ROI mechanisms that efficiently boost the quality of the ROI region while maintaining average quality. Furthermore, our ablation study successfully verifies the effectiveness of our $4$ proposed mechanisms in improving ROI quality.

\appendices
\section{Motivating Application Scenarios} \label{app:scenario}

In Section~\ref{sec:ROI-JSCC_Overview}, we assumed the position of ROI in the image is known to the transmitter and receiver. Here, we explain how the ROI position can be obtained as reliably with high accuracy in real scenarios such as autonomous driving and VR/AR. The details are as follows:

[Vehicles] According to a report from the National Highway Traffic Safety Administration (NHTSA), most crashes (more than 100 out of 130 crashes) involving autonomous vehicles are vehicle-to-vehicle collisions. Among these, most of them are secondary collisions, where a following vehicle crashes into a leading vehicle that suddenly brakes due to a sudden obstacle \cite{national2022summary,wang2022temporal}.

In such cases, a possible solution proposed in the literature is for the leading vehicle to transmit its front-view scene to the following vehicle in real time, allowing the latter to have sufficient time for decision-making~\cite{zhang2021multi}. This paper envisions such a scenario, in which fast tasks such as segmentation and collision probability prediction identify ROI tiles containing the sudden obstacle ahead, which are then transmitted through the ROI-JSCC system to minimize transmission latency.

[VR/AR devices] Most VR/AR headsets are equipped with dedicated sensors or algorithms that track the user’s focal point, which corresponds to the ROI patch in our framework~\cite{wei2023preliminary,aziz2024evaluation}. Hence, devices' sensor determines the location of ROI. The hardware provides the position of ROI(focal position), which developers can easily access through the provided SDK to build applications. Thus, we can assume the ROI position is given as the user’s focal point, which can be provided with high accuracy by the current commercial devices. This is another motivating scenario of our work to focus only on JSCC-based ROI-image transmission with the known-ROI position assumption.

\section{Extensive Experiments} \label{app:extensive}

\subsection{Main Results}

\begin{figure*}[t]
    \centering
    \includegraphics[width=0.8\linewidth]{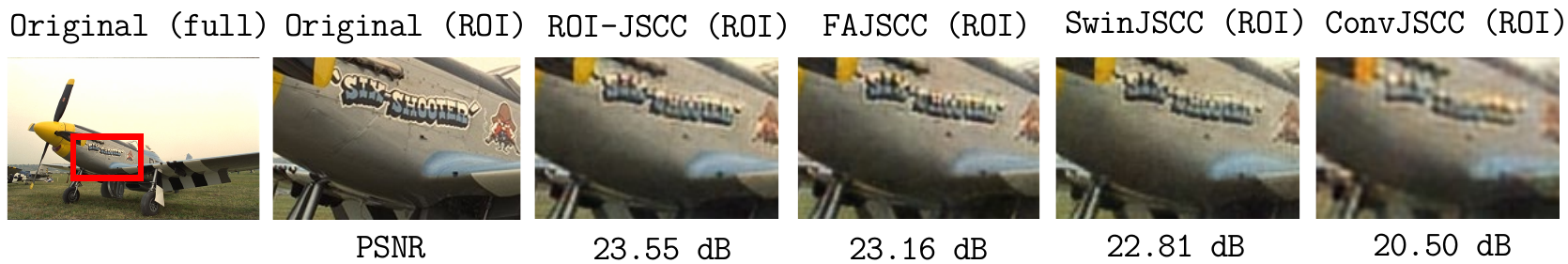}
    \caption{The ROI quality of the reconstructed Kodak image at $\CPP=1/12$, $\SNR=4$dB. The red box is the ROI of the transmitted image.}\label{fig:visual result}
\end{figure*}

\noindent \textbf{Visual Inspection:}
Figure~\ref{fig:visual result} compares ROI visualization results for the reconstructed Kodak image at $\CPP = \frac{1}{12}$ and $\SNR = 4$dB. In this result, our ROI-JSCC shows the clearest reconstruction with the highest $\PSNR_{\ROI}$. For instance, FAJSCC induces color jittering in the text, while SwinJSCC causes blurring near the connection between the wing and the main body. In the case of ConvJSCC, the degradation is so severe that the text becomes barely readable. In contrast, ROI-JSCC effectively reconstructs the image without these artifacts and preserves high ROI quality. In the case of $\PSNR_{\Avg}$, compared to the recent two SOTA models, FAJSCC ($28.01$dB) and SwinJSCC ($27.81$dB), our ROI-JSCC ($27.79$dB) shows minimal degradation.

\begin{figure*}
\centering
\includegraphics[width=0.65\linewidth]{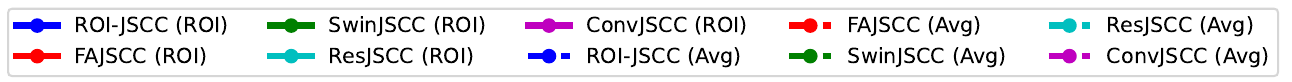}\\
\begin{multicols}{4}
\centering
\includegraphics[width=1.0\linewidth]{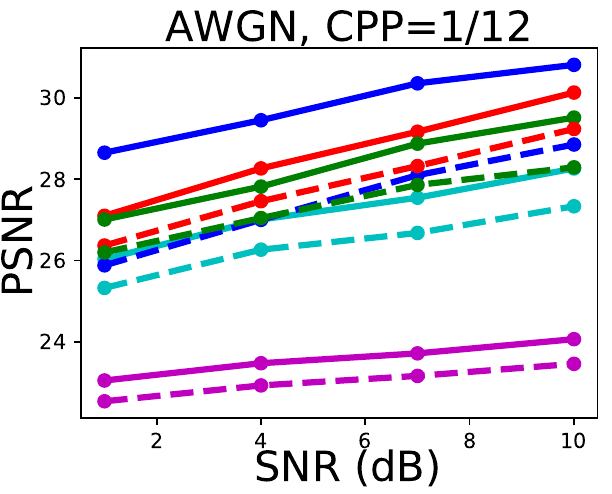}\\

\includegraphics[width=1.0\linewidth]{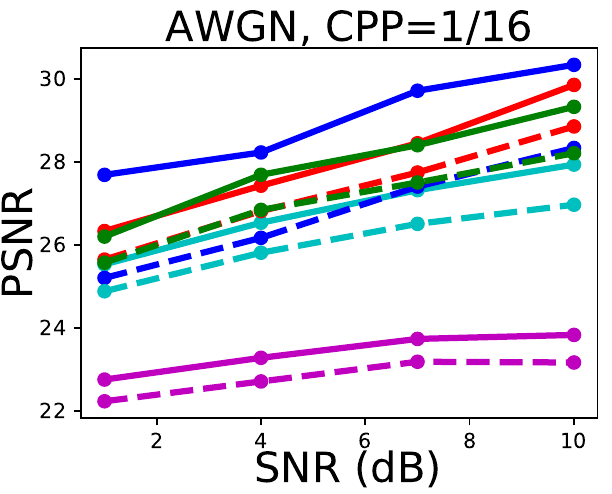}\\

\includegraphics[width=1.0\linewidth]{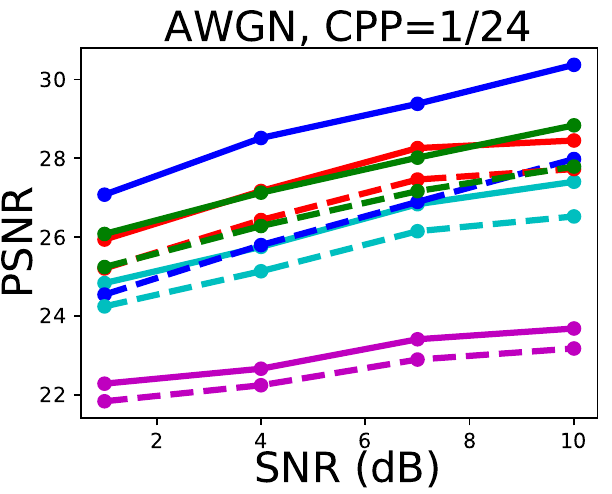}\\

\includegraphics[width=1.0\linewidth]{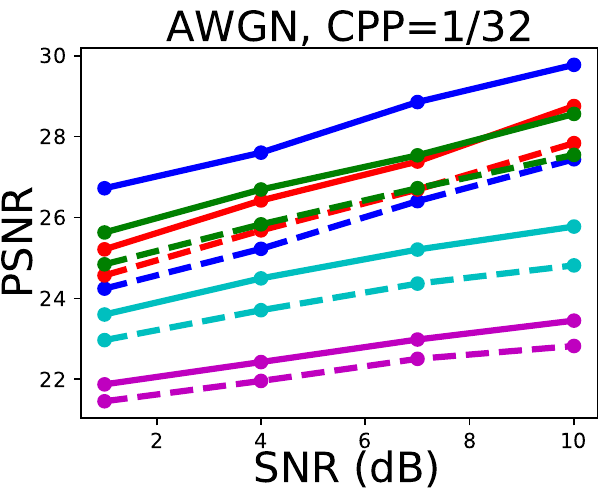}\\
\end{multicols}

\caption{ROI and average PSNR results under different SNR and CPP environments. Different from the main setting, we set $n_h=n_w=8$, i.e., view image as $8 \times 8$ patches.}\label{fig:patch08_psnr_result}
\end{figure*}

\noindent \textbf{ROI Robustness:} Here we discuss the robustness of our ROI-JSCC for various ROI settings. Firstly, without any fine-tuning and retraining, we changed the number of patches to $8 \times 8$, which are different from the training setting of $4 \times 4$. Figures~\ref{fig:patch08_psnr_result} show performance results for this $8 \times 8$ patches settings. Interestingly, compared to $4 \times 4$ patches setting, the ROI performances of our ROI-JSCC show significant improvement while maintaining average performances. This verifies the robustness of our ROI-JSCC for various patch number settings.

This significant performance improvement has two reasons. Firstly, when the number of patches becomes $8 \times 8$, the ratio of ROI patch in the full image becomes much smaller. As a result, it is easy to increase ROI performance, since increasing small ROI areas is easier than large ROI areas. Secondly, as the number of patches increases, the probability that a randomly selected patch has only simple patterns becomes higher. When patches are divided into $4 \times 4$, the randomly selected patch will have some part of the complex patterns with high probability. Complex patterns, such as animals or statues, are difficult to reconstruct. On the contrary, when patches are divided into $8 \times 8$, the probability that a randomly selected patch has complex patterns becomes lower. As a result, more patches will have only simple patterns. Simple patterns, such as sky-like backgrounds, are easy to reconstruct.

In addition to the above discussion for various patch number settings, we discuss various aspects of the ROI Robustness of our ROI-JSCC as follows:

\begin{itemize}
    \item Variants of ROI Positions: Our experiments are conducted by selecting ROI positions randomly (but same random seed as the other models for fair comparison). Our results show that our ROI-JSCC always shows outstanding ROI performances for different metrics (PSNR and SSIM) under various communication environments (various CPP and SNR settings of AWGN, and fast Rayleigh fading channels). Thus, our ROI-JSCC is robust under varying ROI-positions. 

    \item Resolution Variants of ROI Patches: Note that the width and height sizes of DIV2K, and Kodak datasets are varying from $1356$ to $2040$, and from $512$ to $768$ respectively. Moreover, as the number of patches changes from $4 \times 4$ to $8 \times 8$, the resolution of ROI patches also becomes different. We verified the performance of our ROI-JSCC in these settings. Thus, our ROI-JSCC is robust under resolution variants of ROI patches. 
        
    \item Multiple ROIs: Regarding multiple ROIs, we acknowledge that supporting them is necessary for certain applications; however, it represents a nontrivial extension of the current framework. For instance, when considering two ROIs, the model must be trained for all possible ROI pairs, leading to a training complexity proportional to the square of the number of patches. Therefore, directly extending the current ROI-JSCC to handle multiple ROIs is not straightforward. We leave this as a direction for future work.

\end{itemize}

As a result, our ROI-JSCC is robust for various $n_h \times n_w$ patch settings, ROI positions, and resolution of ROI areas. Although we leave multiple ROI issues as future work, our ROI-JSCC is already able to give useful help for existing applications, especially for VR/AR and collision avoidance of autonomous vehicles that are discussed in Appendix~\ref{app:scenario}. Note that, in VR/AR settings, we discussed setting humans' foveal areas as ROI areas to increase user experience. By the physiology of the eye, a usual human has only one foveal area~\cite {levin2024adler}. Thus, it is sufficient to only consider single ROI areas in VR/AR applications. 

In the case of collision avoidance of autonomous vehicles, supporting only one ROI region may be sufficient. The probability that multiple ROIs are needed to avoid collisions seems low, since it is less likely for several collisions to occur simultaneously. For example, it is difficult to imagine a situation where multiple collision risks arise simultaneously—an object suddenly appears in front of the driver while vehicles in the opposite lanes change toward the driver at the same time.\\

\subsection{Comparison with Traditional Approach}
In the early 2000s, several studies explored JSCC for regions of interest (ROI). However, these approaches are not true JSCC methods and not suitable for the scenarios considered in our work. Using the classical ROI–JPEG2000–based JSCC as an example, we provide a detailed analysis as follows:

\begin{figure*}[t]
\centering
\includegraphics[width=1.0\linewidth]{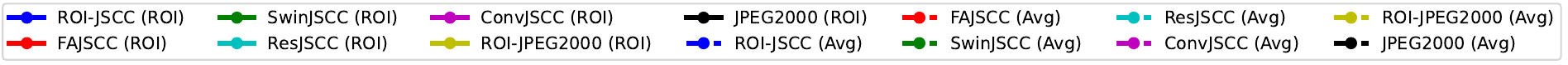}\\
\begin{multicols}{3}
    \centering
    \includegraphics[width=0.8\linewidth]{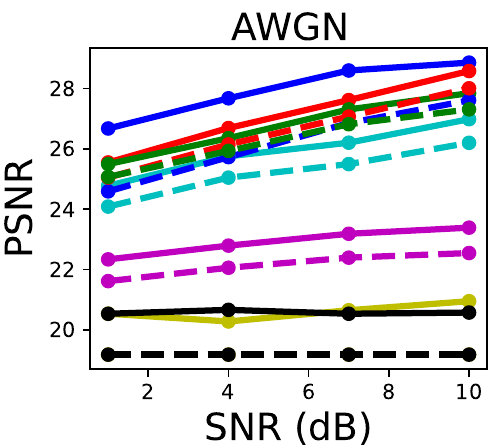}

    {\small ~~~~~~$128 \times 128$ resolution}

    \includegraphics[width=0.8\linewidth]{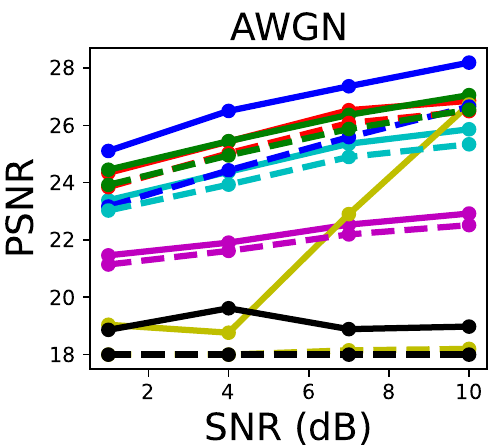}
    
    {\small ~~~~~~$256 \times 256$ resolution}

    \includegraphics[width=0.8\linewidth]{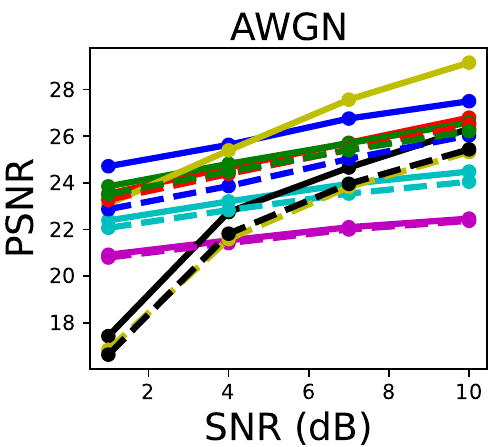}
    
    {\small ~~~~~~$512 \times 512$ resolution}
\end{multicols}
\caption{Comparison with deepJSCC methods and JPEG2000s under various image resolutions, and SNRs. The image resolutions are depicted below each figure. 
}\label{fig:JPEG2000_with_deppJSCCs}
\end{figure*}

\begin{enumerate}
    \item[(1)] Likewise, other ROI-JSCC of the early 2000s~\cite{zhang2006roi,sun2006region}, the ROI-JPEG2000-based JSCC is not considered a true JSCC in the modern sense. It relies on separate source coding (JPEG2000) and channel coding (e.g., LDPC) schemes, with only weak joint optimization between the JPEG2000 and channel coding (LDPC) parameters.

    \item[(2)] Based on our implementation of ROI-JPEG2000, its applicability is limited and not aligned with wireless communication and with our target scenarios, such as low-latency scenarios (e.g., autonomous vehicles) and/or computationally efficient devices (e.g., AR/VR).
    
    Figure~\ref{fig:JPEG2000_with_deppJSCCs} is our implementation of ROI-JPEG2000 and JPEG2000, assuming the use of optimal channel codes (i.e., hypothetical capacity-achieving codes). JPEG2000 is implemented by the open source library OpenJPEG, and ROI-JPEG2000 by designing an ROI-based bits-usage per pixel (BPP) allocation mechanism for JPEG2000. The ROI-based BPP allocation mechanism is a modified version of our ROI-JSCC's bandwidth allocation. The detailed implementations are in Appendix~\ref{app:JPEG2000}, and~\ref{app:ROIJPEG2000}. Then, two observations can be made.

    \begin{itemize}
        \item (ROI-)JPEG2000 codes have a dedicated structure to a specific communication scenario (e.g., SNR, size of images, etc), which means that the encoding and decoding are not universally good. As seen in Figure~\ref{fig:JPEG2000_with_deppJSCCs}, some (ROI-)JPEG2000 results (black: JPEG2000, yellow: ROI-JPEG2000) exhibit flat behavior over varying SNRs. This occurs when the channel capacity is below the minimum bitstream size required by (ROI-)JPEG2000. This is because of its coding structure, (ROI-)JPEG2000 always requires a certain number of bits to store code's metadata (headers, codeblock information, quantization parameters, packet markers, etc.), even for simple images~\cite{taubman2002}. Consequently, when the channel capacity falls below this threshold, (ROI-)JPEG2000 encodes the image at its lowest possible rate, exceeding the intended BPP constraint. Therefore, (ROI-)JPEG2000-based JSCC is only applicable when the channel capacity is large enough or the image resolution is high enough for the metadata overhead to remain relatively small. To solve this, the entire code structure must be redesigned for each environment, e.g., SNRs, channels, image sizes, etc, which requires huge extra efforts and hardware complexities.

        \item Another major drawback of (ROI-)JPEG2000-based JSCC, perhaps even more critical, lies in its computational latency. Since (ROI-)JPEG2000 involves iterative optimization of wavelet parameters, it requires approximately 10 seconds (Xeon(R) CPU E5-2650 v3 used) for encoding and decoding (even without channel coding) of DIV2K high-resolution images on our machine, whereas the proposed ROI-JSCC is much more efficient, taking only about 0.5 seconds (P100 GPU used). Even on the best-performing high-end server computer reported in the literature, JPEG2000 still requires around 60 ms only for encoding~\cite{fastcompression}. In contrast, on the edge device, our ROI-JSCC is expected to operate in $\sim 20$ ms for end-to-end encoding and decoding.\footnote{Inference of YOLOv8s (29.7 GFLOPs) with on-device GPU takes in 7.94 ms~\cite{latency_jetson}. For 2K images, our ROI-JSCC uses $\sim 73$GFLOPs.} Considering hardware performances, the actual latency gap will be much larger when ROI-JPEG and ROI-JSCC are implemented on the same device. Furthermore, because the coding architecture of JPEG2000 is inherently incompatible with parallel processing, its potential for computational acceleration on modern hardware remains limited.

    \end{itemize}

\end{enumerate}

\section{Future Research} \label{app:future}
In this research, we proposed the first ROI-JSCC framework with a detailed model architecture and optimization methods. Moreover, we extensively verified our ROI-JSCC performances under various performance metrics, channels, and ROI settings. Based on our work, appropriate future works are as follows:

\begin{itemize}
    \item \textbf{Multimodal ROI-JSCC for Autonomous vehicles:} In reality, autonomous vehicles aggregate different types of data from cameras, radars, LiDARs. Transmitting these various data with ROI guidance will be much efficient than our current ROI-JSCC. Moreover, in the case of a collision avoidance scenario, our $\log_2(n_h n_w)$ bit length ROI position information also contains some part of the information for collision probability. Finding how to use such an ROI position for the receiver's autonomous driving system is also an interesting research direction.
 
    \item \textbf{Implementing ROI-JSCC on VR/AR devices:} Current eye tracking technologies provide fast and reliable foveal area~\cite{krafka2016eye,valliappan2020accelerating,chen2025ex}, and using this foveal area as ROI for our ROI-JSCC is sufficient to increase the user experience as we discussed in Appendix~\ref{app:scenario}. However, integrating eye tracking technologies and our ROI-JSCC can give a synergistic effect to each other. For example, neural features of ROI-JSCC can help estimate the next gaze point trajectory, and the ROI feature processing of our ROI-JSCC can be faster by using values obtained from eye-tracking estimation, rather than processing ROI embedding blocks.

\end{itemize}

\section{JPEG2000 with Optimal Channel coding}
\label{app:JPEG2000}

Like JSCC uses CPP as an available bandwidth resource constraint, source coding uses bits-per-pixel (BPP) as a data resource constraint. By the information theory, the BPP $R$ for reliable data transmission with channel capacity achieving code is as follows\cite{bourtsoulatze2019deep}:
\begin{align*}
    nR \le k C_{channel},
\end{align*}
where $n:=H \times W \times 3$ is the number of pixels in image $\mathbf{x}$, $k$ is the number of transmitted symbols, and $C_{channel}$ is the channel capacity, which is specified by the channel properties. 
By dividing $n$ in the both sides, the maximum possible BPP $R_{\MAXX}$ is as follows:
\begin{align*}
    R_{\MAXX} = \CPP \times C_{channel}.
\end{align*}
Thus, $R_{\MAXX}$ can be calculated via the given CPP and $C_{channel}$. Note that $C_{channel}=\log_2(1+\SNR)$ for the complex Gaussian channels that we consider. Since $C_{channel}$ can be achieved via ideal capacity-achieving channel coding and modulation methods, the actually reliably transmitted BPP is lower than $R_{\MAXX}$. Thus, our JPEG2000 with $R_{\MAXX}$ should be one of the upper bounds of traditional coding methods. We implemented JPEG2000 with the open source library OpenJPEG.

\section{ROI-JPEG2000}
\label{app:ROIJPEG2000}

Motivated by our ROI-guided bandwidth allocation, we make a new baseline ROI-JPEG2000 to compare with our ROI-JSCC. ROI-JPEG2000 divides image patches based on the ROI map.\footnote{Remind the ROI map explained in Section~\ref{sec:ROI-JSCC_Overview}.} Then, likewise ROI-based bandwidth allocation proposed in the main paper, we propose ROI-based BPP allocation as follows:
\begin{align*}
    R_{\ROI} =& (1 + \eta \tau) R_{\MAXX},\\
    R_{\ROP} =& R_{\MAXX},\\
    R_{\RONI} =& (1 - \eta \tau) R_{\MAXX},
\end{align*}
where $\eta$ and $\tau$ are same with the our ROI-JSCC. Likewise ROI-based bandwidth allocation, this ROI-based BPP allocation enhances the ROI quality with only a minimal degradation in average performance.

\end{document}